\documentstyle[twoside,12pt]{article}
\newtheorem{theorem}{Theorem}

\newtheorem{corollary}{Corollary}
\newtheorem{definition}{Definition}

\author{
\normalsize {\bf B. Shklyar} \\
\normalsize Dept. Math.\&Comp. Sci. \\
\normalsize Bar-Ilan University \\
\normalsize Ramat-Gan 52900 \\
\normalsize Israel}
\title{\normalsize {\bf APPROXIMATE CONTROLABILITY BY CONTROL CONSTRAINTS
FOR INFINITE DIMENSIONAL SYSTEMS}}
\date{}

\begin{document}

\maketitle

\begin{abstract}
{For linear infinite systems the approximate controllability problem by
control constraints is considered. Controllability conditions represented
via system parameters are obtained.}

Partial differential control systems and control systems with delays are
considered as an example.
\end{abstract}

\section{Problem statement}

Let $X,Y,U$, be complex Banach spaces. Consider the abstract evolution
equation

\begin{equation}  \label{e1}
\dot{x}(t)=Ax(t)+Bu(t),
\end{equation}

\begin{equation}
x(0)=x_0,  \label{e2}
\end{equation}
where $x(t)\in X$ is a current state, $x_0\in X$ is an initial state; 
$u(t)\in U,u(.)\in L_2([0,t_1],U)$ is a control; $A:X\to X$ is a linear
unbounded closed operator whose domain $D(A)$ is dense in $X;\ B:U\to X$ is
a linear bounded operator.

We assume the problem (\ref{e1})-(\ref{e2}) to be uniformly well-posed \cite
{Hl-Phlps57}. It follows from this assumption that $A$ generates a strongly
continuous semigroup $S(t)$ on $X$ in the class $C_0$ \cite{Hl-Phlps57}. We
consider only weak solutions \cite{Hl-Phlps57} of the above equation.

As usual $X^{*}$ is a dual space, $A^{*}$ denotes an adjoint operator for
the operator $A$. If $x\in X$ and $f\in X^{*}$, we will write $(x,f)$
instead of $f(x)$.

For any set $K\subset X$ we denote by $\overline{K}$ the closure of $K$ with
respect to the norm of $X$.

Let $K^{\perp }$ be the set $\{y\in X^{*}:(x,y)=0,\forall x\in K\}$.

As usual we denote by ${I\!\!R}$ the set of real numbers and by 
$I\!\!R^n$ the $n$-dimensional vector space.

We assume in the sequel that $u(t)\in \Omega , t\ge 0$ , where $\Omega $ is 
a closed convex cone.

The attainable set $K(t)$ for equation (\ref{e1})-(\ref{e2}) is defined by
the formula:

\begin{equation}
K(t)=\{x\in X:\exists u(.)\in L_2([0,t_1],U),x=x(t_1)\},  \label{e3}
\end{equation}
where $x(t),x(0)=0$ is a weak solution of the equation (\ref{e1}) 
corresponding to the control $u(\cdot )$.

Together with the set (\ref{e3}) we will use the set

\begin{equation}
K_\Omega (t)=\{x\in X:\exists u(.)\in L_2([0,t_1],\Omega ),x=x(t_1)\}.
\label{e4}
\end{equation}

We assume $A$ to have the properties:

(i) The domain $D(A^{*})$ of the operator $A^{*}$ is dense in $X^{*}$.

(ii) The operator $A$ has a purely point spectrum $\sigma $ which is either
finite or has no finite limit points and each $\lambda \in \sigma $ has a
finite multiplicity.

(iii) Let the numbers $\lambda _i\in \sigma ,i=1,2,\ldots $ be enumerated in 
the order of non decreasing real parts,  let $\alpha _i$ be a multiplicity of 
$\lambda _i\in \sigma $, let $\varphi _{ij},i=1,2,\ldots ,j=1,2,\ldots ,
\beta _i, \beta _i\le \alpha _i,A\varphi _{i\beta _i}=\lambda _i\varphi _
{i\beta _i}$ be generalized eigenvectors of the operator $A$, and
let  $\psi_{kl}, k=1,2,\ldots , l=1,2,\ldots ,\beta _k$, be 
generalized eigenvectors of the adjoint operator $A^{*}$, such that

\[
(\varphi _{p\beta _p-l+1,}\psi _{jk})=\delta _{pj}\delta
_{lk}, \ p,j=1,2,\ldots , \ l=1,\ldots ,\beta _p, \ k=1,\ldots ,\beta _j. 
\]
We suppose that $\lim_{i\rightarrow \infty }{\rm Re}\lambda _i=-\infty $, 
and there exists a moment $T,T\ge 0$ such that for each $x\in X, \alpha \in 
I\!\!R$ 
\begin{equation}
S(t)x=\sum_{j=0}^{N_a}\exp (\lambda _jt)\sum_{l=0}^{\beta _j-1}\frac{t^l}{l!}
\sum_{k=l+1}^{\beta _j}\varphi _{jk-l}(x,\psi _{jk})+O(\exp (\alpha t)),
\label{e4.1}
\end{equation}
where $N_\alpha $ is a natural number such that 
${\rm Re}\lambda _j<\alpha ,$ $j=1,2,...,N_\alpha $.

(iv) If $x\in X,g\in X^{*}$ and $(S(t)x,g)\equiv O(\exp (\alpha t))$ for
any $\alpha \in I\!\!R$ then $(S(t)x,g) \equiv 0$ for each $t>T$.

The weak solution $x(t)$ of the equation (\ref{e1})-(\ref{e2}) is evaluated by
the following variation of parameters formula \cite{Hl-Phlps57}: 
\begin{equation}
x(t)=S(t)x_0+\int_0^tS(t-\tau )Bu(\tau )d\tau .  \label{e5}
\end{equation}

\begin{definition}
The equation (\ref{e1}) is said to be approximately $\Omega $-controllable, if
for any $x_0,x_1\in X$ and $\varepsilon >0$ there exists 
$u(\cdot)\in L_2^{{\rm loc}}([0,+\infty ),\Omega )$ and a moment 
$t_1,0<t_1<+\infty $, for which the corresponding solution
$x(t),x(0)=0$ of the equation (\ref{e1}) is such that 
$\Vert x_1-x(t_1)\Vert <\varepsilon $. 
\label{D1}
\end{definition}

\section{Main results}

Let $\Omega \subseteq U$ be a closed convex cone.


\begin{definition}
The linear functional $g_1\in U^{*}$ is said to be not greater then the 
linear functional $g_2$ $\in U^{*}$ with respect to $\Omega $, if 
\begin{equation}
(u,g_1)\leq (u,g_2),\forall u\in \Omega . \label{e6}
\end{equation}
We will denote the inequality (\ref{e6}) by 
\begin{equation}
g_1\leq _\Omega g_2.  \label{e7}
\end{equation}
\end{definition}

\begin{theorem}
\label{T1}The equation (\ref{e1}) is approximately $\Omega $-controllable, if
and only if the inequality 
\begin{equation}
B^{*}S^{*}(t)g\leq _\Omega 0,{\rm \ a.e. \ on \ } [0,+\infty )  \label{e11}
\end{equation}
implies 
\[
g=0.
\]
\end{theorem}

{\bf Proof.} {\it Sufficiency.} Assume the equation (\ref{e1}) be not
approximately $\Omega $-controllable, i.e. $\overline{K_\Omega }\neq X$ . 
As $K_\Omega $ is a convex cone also, the origin is a boundary point of 
$K_\Omega $ \cite{Blkrsh76}, so either there exists a plane of
support containing the origin or for each $\epsilon >0$ there exists
an element $x_\epsilon \in \overline{K_\Omega }\backslash {\rm int}
K_\Omega ,\|x\|<\epsilon $, contained in a plane of support for $K_\Omega $. 
It means that for any $\epsilon >0$ there exists $g\in X^{*},g\neq 0$, 
such that 
\begin{equation}
(x,g)\leq \varepsilon ,\forall x\in K_\Omega (t_1),\forall t_1>0.
\label{e12}
\end{equation}
Using (\ref{e5}) with $x_0=0$ in (\ref{e4}) and (\ref{e4}) in (\ref{e12}) 
we obtain 
\begin{eqnarray}
\int\limits_0^{t_1}(S(t_1-\tau )Bu(\tau ),g)d\tau &<&\varepsilon 
\label{e13} \\
\forall t_1>0, \ \forall u(\cdot ) &\in &L_2([0,t_1],\Omega )  
\nonumber
\end{eqnarray}

Let there exist $u_0\in \Omega \ $and $t_1>0$ such that the set 
\[
\Delta=\{t\in [0,t_1]:(S(t_1-\tau )Bu_0,g)>0\}, 
\]
has a positive measure. Let

\begin{equation}
u^0(t)=\left\{ 
\begin{array}{c}
0,{\rm if \ }t\notin \Delta, \\ 
Lu_0,{\rm if \ }t\in \Delta,
\end{array}
\right.  \label{e14}
\end{equation}
where $u_0\in \Omega ,L>0$. Obviously, $u^0(\cdot )\in L_2([0,t_1],\Omega
),\forall L>0$. Substituting (\ref{e14}) to (\ref{e13}), using the abstract
integration by parts and Euler-Lagrange Lemma \cite{Yang69} we obtain 
\begin{equation}
L\int\limits_{\Delta }T(S(t_1-\tau )Bu_0,g),g)d\tau <\varepsilon ,\forall L>0.
\label{e15}
\end{equation}
However the inequality (\ref{e15}) cannot be true for any $L>0$, if the 
measure of $\Delta $ is positive. Hence the inequality (\ref{e11}) holds with 
$g\neq 0$ for a.e. $t\in [0,t_1]$ and for arbitrary $u\in \Omega $, and this
contradicts to the condition of the theorem. This proves the sufficiency.

{\it Necessity. }Assume that the inequality (\ref{e11}) holds for some $%
g\neq 0$, for a.e. $t\in [0,+\infty )$. As shown above, this inequality is
equivalent to the inequality (\ref{e12}) which holds with $g\neq 0,$ that is 
$\overline{K_\Omega }\neq X$. This completes the proof of the theorem. $\Box 
$

There exists a lot of equations (\ref{e1}) such that the operator $S(t)$ is
injective for all $t\geq 0$. However in the general case it is impossible to
assure that $S(t)$ is necessarily injective for each $t\geq 0$; in this case
there exists $\zeta >0$ and $x\in X,$ $x\neq 0$, such that $S(t)x=0,$ $t\geq
\zeta $, and the same is true for the operator $S^{*}(t)$. Let $h_g=\min
\{t:t\geq 0,$ $S^{*}(t)g=0\}$. Obviously, $h_0=0;h_g$ $>0$ for each 
$g\neq 0$. We assume $h_g=+\infty $ if $S^{*}(t)g\neq 0$ for any $t\geq 0$. 
If $S^{*}(t)$ is injective for any $t\geq 0$, then we have $h_g=+\infty $ 
for any $g\neq 0$.

Systems with delays in an argument provide a number of  non-trivial examples
of operators $S(t)$, where $S^*(t)$ are not injective for some $h>0$.

The following result obtained by means of the Theorem \ref{T1} provides an
approximate $\Omega $-controllability criterion, represented via parameters
of the equation (\ref{e1}).

\begin{theorem}
\label{T2}For the equation (\ref{e1}) to be approximately $\Omega $
-controllable, it is necessary and sufficient, that $\qquad $
\end{theorem}

\begin{enumerate}
\item  
\begin{equation}
\overline{{\rm range}}\{\lambda I-A,B\}=X,\quad \forall \lambda \in \sigma ;
\label{e17}
\end{equation}

\item  the conditions 
\begin{eqnarray}
S^{*}(h_g)g=0,  \label{e18} \\
B^{*}S^{*}(\tau )g\leq _\Omega 0,0\leq \tau \leq h_g<+\infty   \nonumber
\end{eqnarray}
hold if and only if $g=0$;

\item  the operator $A^{*}$ has no real eigenvector $\eta $ such that 
\begin{equation}
B^{*}\eta \leq _\Omega 0.  \label{e19}
\end{equation}
\end{enumerate}

{\bf Proof. }{\it Sufficiency. }Let the conditions (\ref{e11}) and  
(\ref{e17})--(\ref{e18}) hold. We will prove that if $g\neq 0$, then for a
sufficiently large $\alpha $ the set 
\begin{eqnarray*}
J_\alpha &=&\{j:1\leq j\leq N_\alpha ,{\rm \ }\exists {\rm \ }\gamma
_j,1\leq \gamma _j\leq \beta _j\ {\rm such\ that} \\
{\rm \ }(\varphi _{jk},g) &=&0,k=1,...,\gamma _{j-1},(\varphi _{j\gamma
_j},g)\neq 0\}
\end{eqnarray*}
is not empty.

Let 
\begin{equation}
(\varphi _{jk},g)=0, \ j=1,2,..., \ k=1,...,\beta _j.  
\label{e19.1}
\end{equation}

It follows from (\ref{e19.1}) and (\ref{e4.1}) that 
\begin{equation}
(S(t)x,g)=O(\exp (\alpha t)),\forall x\in X,\forall \alpha \in I\!\!R.
\label{e20.1}
\end{equation}
$\ \ $By (iv) and (\ref{e20.1}) we obtain
\begin{equation}
S^{*}(T)g=0.  \label{e20.2}
\end{equation}
Hence $h_g\leq T$. It follows from (\ref{e20.2}) and (\ref{e11}) that 
\begin{equation}
S^{*}(h_g)S^{*}(T-h_g)g=0,  \label{e20.3}
\end{equation}
\begin{equation}
B^{*}S^{*}(\tau )S^{*}(T-h_g)g\leq _\Omega 0,0\leq \tau \leq h_g.
\label{e20.4}
\end{equation}
So we obtain from (\ref{e20.3}), (\ref{e20.4}), and (\ref{e18}) that 
$S^{*}(T-h_g)g=0$. Continuing this process by the similar way, we will 
obtain after a finite number of steps that $g=0$, i.e. we have a
contradiction.

Using $Bu$ instead of $x$ in (\ref{e4.1}) and (\ref{e4.1}) in the 
bilinear form $(S(t)Bu,g)$ , we obtain 
\begin{equation}
\ \ (S(t)Bu,g)=\sum_{j=0}^{N_a}(\exp (\lambda _jt)\sum_{l=0}^{\beta _j-1}%
\frac{t^l}{l!}\sum_{k=l+1}^{\beta _j}(\varphi _{jk-l},g)(Bu,\psi
_{jk})+O(\exp (\alpha t)).  \label{e21}
\end{equation}

Let $J_1=\{j\in J_\alpha :\lambda _j$ is real$\};$ $\ J_2=\{j\in J_\alpha
:\lambda _j$\ is complex$\}$. Apparently, $J_\alpha =J_1\cup J_2,J_1\cap
J_2=\emptyset $.

We use the following notations: 
\begin{eqnarray*}
\mu _1 &=&\max \{\lambda _j,j\in J_1\},I_1=\{j\in J_1,\lambda _j=\mu _1\}, \\
l_1 &=&\max \{\beta _j-\gamma _j,j\in I_1\},I_3=\{j\in I_1,\beta _j-\gamma
_j=l_1\}; \\
\mu _2 &=&\max \{{\rm Re}\lambda _j,j\in J_2\},I_2=\{j\in J_2,{\rm Re}
\lambda _j=\mu _2\}, \\
l_2 &=&\max \{\beta _j-\gamma _j,j\in I_2\},I_4=\{j\in I_2,\beta _j-\gamma
_j=l_2\}.
\end{eqnarray*}

Let $J_1=\emptyset $. Then $J_2\neq \emptyset $ and (\ref{e21}) can be
written as 
\begin{equation}
\ \ (S(t)Bu,g)=\exp (\mu _2t)t^{l_2}\psi (t,u)+O(\exp (\alpha t)),
\label{e22}
\end{equation}
where 
\begin{equation}
\psi (t,u)=2\sum_{j\in I_4}({\rm Re}((\varphi _{j\gamma _j},g)(Bu,\psi
_{j\beta _j}))\cos {\rm Im}\lambda _jt-{\rm Im}((\varphi _{j\gamma
_j},g)(Bu,\psi _{j\beta _j}))\sin {\rm Im}\lambda _jt). 
\label{e22.1}
\end{equation}

Let $k\in I_4$. It follows from (\ref{e17}) and the definition of numbers 
$\gamma _j$ that for each $k\in N_\alpha $ that there exists $u_k$ such that 
\begin{equation}
\left| (\varphi _{k\gamma _k},g)(Bu_v,\psi _{j\beta _k})\right| \neq 0.
\label{e22.2} 
\end{equation}
Moreover, $O(\exp (\alpha t))=O(\exp (\mu _2t)t^{l_2}),$ because $\mu _2\leq
\alpha $. 

All the functions $cos({\rm Im}\lambda _jt), sin({\rm Im}\lambda _jt)$ 
are linearly independent with zero mean values In virtue of (\ref{e22}) and 
the lemma on almost-periodic functions \cite{Kor-Mar-Pod72} there are  a 
sequence $t_v, \lim_{v\to \infty}$ and a number $v_0$ such that for an 
arbitrary $v\ge v_0$ 
\begin{equation}
(S(t_v)Bu_k,g)>0.  
\label{e23}
\end{equation}

If $J_1\neq \emptyset $ and $J_2=\emptyset $, then (\ref{e21}) can be
written as 
\begin{equation}
(S(t)Bu,g)=\exp (\mu _1t)t^{l_1}\psi (t,u)+O(\exp (\alpha t)),  
\label{e24}
\end{equation}
where 
\[
\psi (t,u)=(Bu,\sum\limits_{j\in I_3}(\varphi _{j\gamma _j},g)\psi _{j\beta
_j}). 
\]

Let $\psi _1=\sum_{k\in I_3}(\varphi _{j\gamma _j},g)\psi_{k\beta _k}$. 
The vectors $\psi _{k\beta_k},k\in I_3$ are linearly independent, hence by the 
definition of numbers $\gamma _j$ we obtain $\psi_1\ne 0$, so vector $\psi _1 $
is an eigenvector of the operator $A$, corresponding to the real
eigenvalue $\mu _1$, and in virtue of the third condition of the theorem 
there exists $u_0\in \Omega $ such that $(Bu_0,\psi _1)>0$. We have also
$O(\exp (\alpha t))=O(\exp (\mu _1t)t^{l_1}),$ because $\mu _1\leq
\alpha $.  Hence
\begin{equation}
\ \ (S(t)Bu_0,g)=\exp (\mu _1t)t^{l_1}(Bu_0,\psi _1)+O(\exp (\mu _1t)t^{l_1}),
\nonumber
\end{equation}
Since $(Bu_0,\psi _1)>0$, there exists $T_1>0$ such that for arbitrary 
$t\geq T_1$
\begin{equation}
(S(t)Bu_0,g)>0.  
\label{e26}
\end{equation}

If $J_1\neq \emptyset $ and $J_2\neq \emptyset ,$ then arguing as in above
cases we can write 
\begin{eqnarray}
\ \ (S(t)Bu_0,g) &=&\exp (\mu _1t)t^{l_1}(Bu_0,\psi _1)+\exp (\mu
t)t^{l_3}\psi (t,u_0)+  \label{e27} \\
&&\ \ O(\exp (\mu _1t)t^{l_1})+O(\exp (\mu _3t)t^{l_3}),  \nonumber
\end{eqnarray}
where $(Bu_0,\psi _1)>0$; 
\[
\psi (t,u_0)=2\sum_{j\in I_5}\alpha _{1j}\cos \delta _jt+\alpha _{2j}\sin
\delta _jt, 
\]
where $\mu _3\leq \mu _2,l\leq l_2,I_5\subseteq I_4;\delta _k\neq \delta
_p,k,p\in I_5$ and there exists $v\in I_5$ such that $\left| a_{1v}\right|
+\left| \alpha _{2v}\right| \neq 0$. Hence by the lemma on almost-periodic
functions there exist a sequence $t_v,\lim\limits_{v\to \infty }t_v=\infty $
and a number $v_1$, such that 
$\exp (\mu t_v)t^{l_3}\psi (t_v,u_0)+O(\exp (\mu _3t_v)t_v^{l_3})>0$ 
for any $v\geq v_0$. If $v\geq v_1$ such that $t_v\geq T_1$, then for the 
sufficiently small $\eta $
\begin{equation}
(S(t)Bu_0,g)>0,t_v\leq t\leq t_v+\eta .  
\label{e28}
\end{equation}

The formula (\ref{e28}) shows that if $g\neq 0$ then (\ref{e11}) doesn't hold 
so the sufficiency follows by Theorem \ref{T1}.

{\it Necessity. }If (\ref{e17}) doesn't hold, then there exists $\lambda \in
\sigma $ and eigenvector $g_\lambda $ of the operator $A^{*}$ such that 
$B^{*}g_\lambda =0$. Therefore $B^{*}S^{*}(t)g_\lambda =\exp (\lambda
t)B^{*}g_\lambda =0,\forall t\geq 0,g_\lambda \neq 0$. This contradicts to
the Theorem \ref{T1}.

Let there exists a vector $g\in X^{*},g\neq 0$, such that $h_g<+\infty $ and 
\begin{equation}
S^{*}(t)g=0,\forall t\geq h_g,  \label{e30}
\end{equation}
\[
B^{*}S^{*}(\tau )g\leq _\Omega 0,0\leq \tau \leq h_g. 
\]
Relations (\ref{e30}) imply 
\[
B^{*}S^{*}(t)g\leq _\Omega 0,\forall t\geq 0, 
\]
where $g\neq 0$. This contradicts to the Theorem \ref{T1}.

If there exists a real eigenvalue $\lambda $ and a corresponding real
eigenvector $\eta _\lambda $ such that $B^{*}\eta _\lambda \leq _\Omega 0$,
we obtain from the last inequality, that $B^{*}S^{*}(t)\eta _\lambda =\exp
(\lambda t)B^{*}\eta _\lambda \leq _\Omega 0,\forall t\geq 0,\eta _\lambda
\neq 0$.

The theorem follows by the contradiction to the Theorem \ref{T1}. $\Box $

\begin{corollary}
\label{C1}Let $b\in U,b\neq 0$. Consider the cone $\Omega =bI\!\!R%
^{+}=\{u\in U:u=b\alpha ,\alpha \geq 0\}$. For the equation (\ref{e1}) to be
approximately $\Omega $-controllable, it is necessary and sufficient that

\begin{enumerate}
\item  The condition (\ref{e17}) holds.
\item  The conditions (\ref{e18}) hold if and only if $g=0$.
\item  The operator $A$ has no real eigenvalues.
\end{enumerate}
\end{corollary}

{\bf Proof. }If $\Omega =bI\!\!R^{+}$, then the condition 2 of the
Corollary is equivalent to the condition (\ref{e19}) of the Theorem \ref{T2}.

\section{Examples}

{\it Example 1. }Let $H$ be a Hilbert space. Consider the equation (\ref{e1})
with the self-adjoint operator $A$ generating a $C_0$-semigroup. It is well-known 
that the spectrum $\sigma $ of $A$ is the sequence $\{\lambda _j,j=1,2,\ldots \}$ 
of real negative numbers; $\lim_{j\to \infty }\lambda _j=-\infty $; the operator 
$A$ has the properties (i)-(iv) with $T=0$. 

Let $\varphi _j,j=1,2,\ldots $be eigenfunctions of $A$ corresponding to
eigenvalues $\lambda _j,j=1,2,\ldots $, and let $I_j=\{k:\lambda
_k=\lambda _j\}$. It is well-known, that the sequence 
$\{\varphi_j,j=1,2,\ldots \}$ 
is complete. One can prove, that for a self-adjoint operator $A$ the condition 
(\ref{e17}) is equivalent to the condition

\begin{equation}
{\rm Ker}\{\lambda I-A\}\bigcap {\rm Ker}\{B^{*}\}=\{0\}.  \label{e31}
\end{equation}
It is easily to show that the condition (\ref{e31}) is equivalent to the 
linear independence of vectors 
\begin{equation}
B^{*}\varphi _k,k\in I_j,j=1,2,...\ .  \label{32}
\end{equation}

Since the sequence of eigenfunctions of $A$ is complete, we obtain that the
operator $S^{*}(t_1)\ $is injective for an arbitrary $t_1\geq 0$, therefore 
the conditions (\ref{e18}) hold for each self-adjoint operator $A$. As all
the eigenvalues of the operator $A$ are real, the condition (\ref{e19}) is
equivalent to the following:

(v) there is no natural $j$ such that $B^{*}\varphi _j\leq _\Omega 0.$

Thus, we obtain the validity of the following theorem:

\begin{theorem}
\label{T3}Let $A$ be a self-adjoint operator in a Hilbert space $H$. The
equation (\ref{e1}) is approximately $\Omega $-controllable if and only if 
(\ref{e31}) and {\rm (v) }hold.
\end{theorem}

Obviously, (v) does not hold, if 
$U=I\!\!R,$ $\Omega =I\!\!R^{+}=\{\alpha \in {I\!\!R,}\alpha \geq 0\}$, 
so the next corollary 
follows.

\begin{corollary}
\label{C2}Let $A$ be a self-adjoint operator in the Hilbert space $H$.
If $U=I\!\!R,$ $\Omega =I\!\!R^{+}$, then the equation (\ref{e1}) 
is not approximately $\Omega $-controllable.
\end{corollary}

To illustrate above theorem consider the following simple example.

Consider the control system described by one-dimensional heat equation 
\begin{equation}
{\frac{\partial x}{\partial t}}(t,\xi )={\frac{\partial ^2x}{\partial \xi ^2}
}(t,\xi )+B(\xi )u(t);  
\label{e30.1}
\end{equation}
\begin{equation}
x(0,\xi )=\varphi (\xi ),0\le \xi \le \pi ,  
\label{e30.2}
\end{equation}
\begin{equation}
x(t,0)=0,x(t,\pi )=0,0\le t<+\infty ,  
\label{e30.3}
\end{equation}
where\ $x,\varphi \in I\!\!R,B(\xi )={\rm col}\{b_1(\xi ),b_2(\xi
)\},\varphi (\cdot ),$ $b_j(\cdot )\in L_2[0,\pi ],j=1,2$.

The equation (\ref{e30.1}) is a particular case of the problem 
(\ref{e1})-(\ref{e3}), where 
\[
X=L_2[0,\pi ], U=I\!\!R^2; (Ax)(\xi )={\frac{d^2x}{d\xi ^2}}(\xi ) 
\]
with the domain 
\[
D(A)=\{x\in C^2[0,\pi ]:x(0)=x(\pi )=0\}. 
\]
It is well-known that the operator $A$ generates a compact self-adjoint 
$C_0$-semigroup; $\sigma =\{-j^2,j=1,2,\ldots \};\alpha _j=\beta _j=1,
\varphi_j(\xi )=\sqrt{2}\sin (j\xi )$ are the eigenfunctions of the operator 
$A$ corresponding to eigenvalues $\lambda _j=-j^2,j=1,2,\ldots $; $A$ is a
self-adjoint operator; for each $\varphi (.)\in X$ the corresponding
solution $x(t,\xi )$ of equation (\ref{e30.1})-(\ref{e30.3}) is expanded
into the series 
\[
x(t,\xi )=
\sum_{j=1}^\infty (\int_0^\pi \varphi (\xi )\sin (j\xi )d\xi)\exp(-j^2t), 
\]
convergent uniformly for any segment $[0,h]$.

Let the closed convex cone $\Omega $ be described by the set 
$\Omega=\{(v_1,v_2)\in I\!\!R^2\}$, where
\begin{eqnarray*}
v_1=c_{11}u_1+c_{12}u_2,\\
v_2=c_{21}u_1+c_{22}u_2, 
\end{eqnarray*}
where  $u_1\ge 0, \ u_2\ge 0$.

The implementation of the Theorem (\ref{T3}) shows the validity of the 
following result:

\begin{theorem}
\label{T4}The equation (\ref{e30.1})-(\ref{e30.3}) is approximately $\Omega $
-controllable if and only if there is no natural $j$ such that
\[
\int\limits_0^\pi (b_1(\xi )c_{11}+b_2(\xi))\sin (j\xi )d\xi\le 0,
\]
and
\[
\int\limits_0^\pi (b_1(\xi )c_{11}+b_2(\xi))\sin (j\xi )d\xi\le 0.
\label{e30.4}
\]
\end{theorem}

{\it Example 2.} Consider the linear hereditary system \cite{Hale77}

\begin{equation}
{\frac d{dt}}x(t)=A_0x(t)+A_1x(t-h)+B_0u(t),  \label{e41}
\end{equation}

\begin{equation}
x(0)=x_0,x(\tau )=\varphi (\tau ),-h\le \tau \le 0,  \label{e42}
\end{equation}
where $x,x_0\in I\!\!R^n,u\in I\!\!R^r,$ $A_0,A_1$ are ${n}\times {n}$ 
constant matrices, $\varphi (.)\in L_2^n[-h,0];B_0$ is a ${n}\times {r}$
constant matrix. It is known that the problem (\ref{e41})-(\ref{e42}) is
well-posed \cite{Hale77}. Hence the problem (\ref{e41})-(\ref{e42}) is a
particular case of the problem (\ref{e1})-(\ref{e2}), where 
\[
X=I\!\!R^n\times L_2^n[-h,0],U=I\!\!R^r; 
\]
the corresponding operator $A$ has the properties (i)-(ii)\cite{Hale77};
the properties (iii)--(iv) hold for $T=nh$ \cite{Blm-Cuk63}, \cite{Hnr70};
the operator $B$ is defined by
$$
Bu=\{B_0u,0\}.
$$ 
Let the closed convex cone $\Omega $ is described by the set 
\[
v=Cu\leq 0,j=1,...,m, 
\]
where $C$ is $r\times m$ constant matrix, $1\leq m\leq r, u\ge 0$.

Using above Theorem \ref{T2}, one can prove the validity of the
following result.

\begin{theorem}
\label{T5}
For approximate $\Omega $-controllability of the system (\ref{e41}) 
it is necessary and sufficient, that

\begin{enumerate}
\item  for any $\lambda \in \sigma $
\begin{equation}
{\rm rank}\{\lambda I-A_0-A_1\exp (\lambda \tau ),B_0\}=n;  
\label{e43}
\end{equation}

\item  the system of equations and inequalities 
\begin{eqnarray}
A_1x=0,  
\label{e44} \\
\ B_0^Tx\leq 0  
\nonumber
\end{eqnarray}
has only the trivial solution;

\item  the system 
\begin{equation}
\stackrel{.}{y}(t)=A_0^Ty(t)+A_1^Ty(t-h)  
\label{e45}
\end{equation}
has no real eigenvectors $g$ such that 
\begin{equation}
g^TB_0C\leq 0.  
\label{e45-1}
\end{equation}
\end{enumerate}
\end{theorem}

{\bf Proof. }The equivalence between (\ref{e43}) and (\ref{e17}) follows
from the results of \cite{Shk91}.

1.Now we have to establish the equivalence between (\ref{e44}) and (\ref{e18}).

1.1. Let $x=\{x_0,\varphi(\cdot )\},g=\{y_0,\psi(\cdot )\}$ be arbitrary
elements from $X$ and $X^*$ correspondingly, and let $F(t)$ be a
fundamental matrix of the system (\ref{e41}).

It follows from the results of \cite{Blm-Cuk63},\cite{Hale77},\cite{Shk85}, 
that
\footnote{The superscript $^T$ denotes a transposition.} 
\begin{equation}
(S(t)x,g)=y^T(t)x_0+\int\limits_0^hy^T(t-\tau )A_1\varphi (\tau -h)d\tau
,\forall t\geq 0,  
\label{e46}
\end{equation}
where 
\begin{equation}
y(t)=F^T(t)y_0+\int\limits_{-h}^0F^T(t+\tau )\psi (\tau )d\tau, t\geq 0.
\label{e47}
\end{equation}
It is easy to show that $y(t)$ is a solution of the equation (\ref{e45}) 
for $t\in [h,+\infty )$. Hence the equality $S^{*}(h)g=0$ is equivalent to 
\begin{equation}
y(h)=0,A_1^Ty(\tau )\stackrel{a.e.}{=}0\ {\rm on\ }[0,h].  
\label{e48}
\end{equation}

Further we are to evaluate $B^{*}S^{*}(\tau )g,0\leq \tau \leq h$. Using in 
(\ref{e46}) $x=Bu,u\in I\!\!R^r$, we obtain 
\begin{equation}
(S(\tau )Bu,g)=y^T(\tau )B_0u,0\leq \tau \leq h,\forall u\in I\!\!R^r.
\label{e49}
\end{equation}
Therefore, the inequality $B^{*}S^{*}(\tau )g\leq 0, \ 0\leq \tau \leq h$ is
equivalent to 
\begin{equation}
B_0^Ty(\tau )\leq 0,0\leq \tau \leq h.  
\label{e50}
\end{equation}
Thus the conditions (\ref{e48}) and (\ref{e50}) are  equivalent to the 
conditions (\ref{e18}).

If (\ref{e44}) holds, then (\ref{e48}) and (\ref{e50}) imply 
\begin{equation}
y(h)=0,y(\tau )\stackrel{a.e.}{=}0\ {\rm on\ }[0,h].  
\label{e51}
\end{equation}

By (\ref{e47}), (\ref{e51}) we obtain 
\begin{equation}
y(\tau )=\exp (A_0\tau )y_0+\int\limits_0^\tau \exp (A_0(\tau -\theta )\psi
(-\theta )d\theta, 0\le \tau \le h  
\label{e52}
\end{equation}

Hence $y(\tau ),0\leq \tau \leq h$ is an absolutely continuous function and
it satisfies the ordinary non-homogeneous differential equation 
\begin{equation}
\dot y(\tau )=A_0y(\tau )+\psi (-\tau ),0\leq \tau \leq h.  
\label{e53}
\end{equation}
The equations (\ref{e51}) and (\ref{e53}) imply 
\[
y_0=0,\psi (\tau )\stackrel{a.e.}{=}0\ {\rm on\ }[-h,0]. 
\]
Therefore, the conditions (\ref{e44}) imply the conditions (\ref{e18}).

1.2. Now we will prove that the conditions (\ref{e18}) imply the condition 
(\ref{e44}).

Let (\ref{e18}) be true and let $x\in I\!\!R^n$  be such that 
\begin{eqnarray}
A_1^Tx &=&0,  \label{e54} \\
\ B_0^Tx &\leq &0  \nonumber
\end{eqnarray}
Consider 
\begin{equation}
y=\{0,x\theta \},-h\leq \theta \leq 0.  
\label{e55}
\end{equation}
Obviously, $y\in X=I\!\!R^n\times L_2^n[-h,0]$. Let $y(t)$ be the
solution of equation (\ref{e45}) on $[h,t]$ with initial condition

\begin{equation}
\xi =\{0,x\cdot (\theta-h)\},0\leq \theta \leq h.  
\label{e56}
\end{equation}

This solution is defined by the formula \cite{Blm-Cuk63}

\begin{equation}
y(t)=\int\limits_h^{2h}F^T(t-\tau )A_1^Tx\cdot (\tau -2h)d\tau ,t\geq h.
\label{e57}
\end{equation}

It follows from (\ref{e57}) that $y(t)\equiv 0,t\geq h$.

Define the function $p(t),0\leq t\leq h$, by the formula:

\begin{equation}
p(-t)=x-A^T_0x\cdot (t-h)\footnote{In accordance with the definition of the function 
$y(t)$ we have 
$\dot{y}(t)\equiv x, 0\le t\le h$.}.  
\label{e58}
\end{equation}

The function $y(t)$ is a solution of the non-homogeneous system

\begin{equation}
\stackrel{.}{y}(t)=A_0^Ty(t)+A_1^Ty(t-h)+q(t),t\geq 0  
\label{e59}
\end{equation}
with the initial condition

\[
y_0=x\cdot (-h),y(\tau )\equiv 0,-h\leq \tau \leq 0, 
\]
where

\[
q(t)=\left\{ 
\begin{array}{c}
0,t>0, \\ 
p(-t),0\leq t\leq h.
\end{array}
\right. 
\]

Therefore \cite{Blm-Cuk63}

\begin{equation}
y(t)=F^T(t)x\cdot (-h)+\int\limits_{-h}^0F^T(t+\tau )p(\tau )d\tau .
\label{e60}
\end{equation}
The formulas (\ref{e46}),(\ref{e60}) and 
$y(t) \ {\rm a.e. }=0 \ {\rm on \ } [h,+\infty)$ 
imply

\begin{equation}
S^{*}(2h)g=0,  
\label{e61}
\end{equation}
where $g=\{x\cdot (-h),p(\cdot )\}$. The condition 
$y(t)\stackrel{a.e.}{=}0\ {\rm on\ }[h,+\infty)$ and  (\ref{e47})  imply 
\begin{equation}
B^{*}S^{*}(\tau )g\leq 0,0\leq \tau \leq 2h,  
\label{e62}
\end{equation}
In account of (\ref{e18}) it follows from (\ref{e61}) and (\ref{e62})
that $g=0$, i.e. $x=0$. Hence, (\ref{e44}) holds, as required.

3. It remains to establish the equivalence between (\ref{e19}) and (\ref{e45-1}). 
We have 
\[
B^{*}g=B_0^Tg_0,\forall g=\{g_0,g(\cdot )\}\in I\!\!R^n\times L_2^n[-h,0]. 
\]
If $\eta =\{\eta _0,\eta (\cdot )\}$ is an eigenvector of the operator $A^*$ 
corresponding to an eigenvalue $\lambda $ then \cite{Hale77} 
\begin{equation}
\eta (\tau )=\eta _0\exp (-\lambda \tau ),-h\leq \tau \leq 0,  \label{e63}
\end{equation}
\begin{equation}
\eta _0^T(\lambda I-A_0-A_1\exp (-\lambda h))=0.  \label{e64}
\end{equation}
Hence, if $\eta=\{\eta_0,\eta(\cdot )\}$ is an eigenvector of the operator 
$A^{*}$ such that (\ref{e19}) holds, than it follows from 
(\ref{e63})-(\ref{e64}) that $C^TB_0^T\eta_0\leq 0$, where $g_0$ is an 
eigenvalue of the system (\ref{e45}), as required. $\Box $

Theorem \ref{T5} is a particular case of the theorem proven in \cite{Shk85}.

\section{Conclusion}

We have obtained the necessary and sufficient conditions for 
the approximate controllability by control constraints for an abstract 
operator differential equation. These results allow us to obtain the 
appropriate controllability conditions for various known classes of 
distributed control systems by the unified manner. Partial
differential control systems and differential-difference control
systems have been considered as an example.

\end{document}